\documentclass[aps,jmp,amsmath,amssymb,reprint]{revtex4-1}
\usepackage{graphicx}
\usepackage{dcolumn,booktabs,array}
\usepackage{bm}
\usepackage{graphicx}
\usepackage{subfigure}
\usepackage{physics}
\usepackage{color}

\begin{document}

\title{Electronic Fabry-P\'erot cavity engineered nanoscale thermoelectric generators}

\author{Swarnadip Mukherjee}
\author{Bhaskaran Muralidharan}%
\email{bm@ee.iitb.ac.in}
\affiliation{Department of Electrical Engineering, Indian Institute of Technology Bombay, Powai, Mumbai 400076, India}



\begin{abstract}
In this work, we aim to design a heterostructure based nanoscale thermoelectric generator that can maximize the waste-heat conversion efficiency at a given output power. The primary objective to be achieved for this is to realize a boxcar-shaped (bandpass) electronic transmission function (R. S. Whitney, Phys. Rev. Lett. 112, 130601 (2014)). In order to achieve that, we propose the use of an electronic analog of optical Fabry-P\'erot cavity over a central resonant tunneling structure. We further explore the optimum design possibilities by varying the geometry of the cavity wall to ensure a nearly perfect bandpass energy filtering of electrons. Based on our findings, we propose a general design guideline to realize such transmission and demonstrate that such devices can be excellent thermoelectric generators compared to the existing proposals in terms of boosting the output power without a cost in efficiency. It is theoretically demonstrated using the non-equilibrium Green's function technique coupled with self-consistent charging effects that an enhancement in the maximum output power up to $116\%$ can be achieved through this scheme at a $10\%$ higher efficiency as compared to resonant tunneling based devices. Furthermore, an elaborate comparative study of the linear response parameters is also presented and explained in terms of the physical transport properties. This study suggests an optimal device design strategy for an improved thermoelectric generator and sets the stage for a new class of thermoelectric generators facilitated via transmission lineshape engineering.

\end{abstract}

\maketitle

\section{Introduction}
Nanostructuring of thermoelectric (TE) materials has acquired unabated precedence over their bulk counterparts \cite{Hicks1992,Hicks1993,Hicks1996,Sofo-Mahan1996,Nakpathomkun2010,Heremans2013,Aniket2015,Majumdar2004} since last two decades due to their highly efficient energy harvesting capability. Over the years, research in this field was primarily focused on achieving high thermoelectric figure-of-merit by means of lineshape engineering \cite{Hicks1992,Hicks1993,Hicks1996,Sofo-Mahan1996,Nakpathomkun2010,Heremans2013}, thermal conductivity reduction through interface engineering \cite{Snyder2008,Harman,Poudel} and enhancement of power factor utilizing energy filtering effects \cite{Bahk,Thesberg,Aniket2017}. The figure of merit concept typically assists in determining whether a material is a good thermoelectric or not. However, when actual device designs are considered, non-linear transport studies \cite{Hershfield,Karbaschi2016} dealing with the trade-off between conversion efficiency and output power of the entire set up \cite{Agarwal2014,Sothmann,Sothmann_Review,Muralidharan2012,Nakpathomkun2010,Esposito1,Esposito3,Bitan2016} have gained precedence. \\
\indent In this context, an important work by R. S. Whitney \cite{Whitney2014,Whitney2015} suggested that in a thermoelectric device set up, a {\it{boxcar}} type electronic transmission function of a particular bandwidth can offer optimum trade-off by maximizing the efficiency at a given power. However, practical design guidelines of such type of devices are not well addressed. Several efforts have been made after that to realize such an electronic transmission feature by proper arrangements of tunnel coupled quantum dots (QD) \cite{Hershfield,Schiegg}.

\indent A few recent studies \cite{Karbaschi2016,pankaj} utilized the miniband feature of superlattice based devices \cite{Broido1995,Tung1996} to achieve the boxcar transmission profile. Further advancing on such ideas, recently, thermoelectric generator (TEG) setups augmented with an electronic anti-reflection cavity (ARC) \cite{pankaj,myTED} have been proposed using the basic thumb rule for ARC design \cite{Pacher2001}.  These ideas proved to be far superior in terms of achieving excellent power-efficiency trade-off in comparison with the competing device proposals 
\cite{Agarwal2014,Sothmann,Karbaschi2016}.
\begin{figure}
	\centering
	\subfigure[]{\includegraphics[scale=0.35]{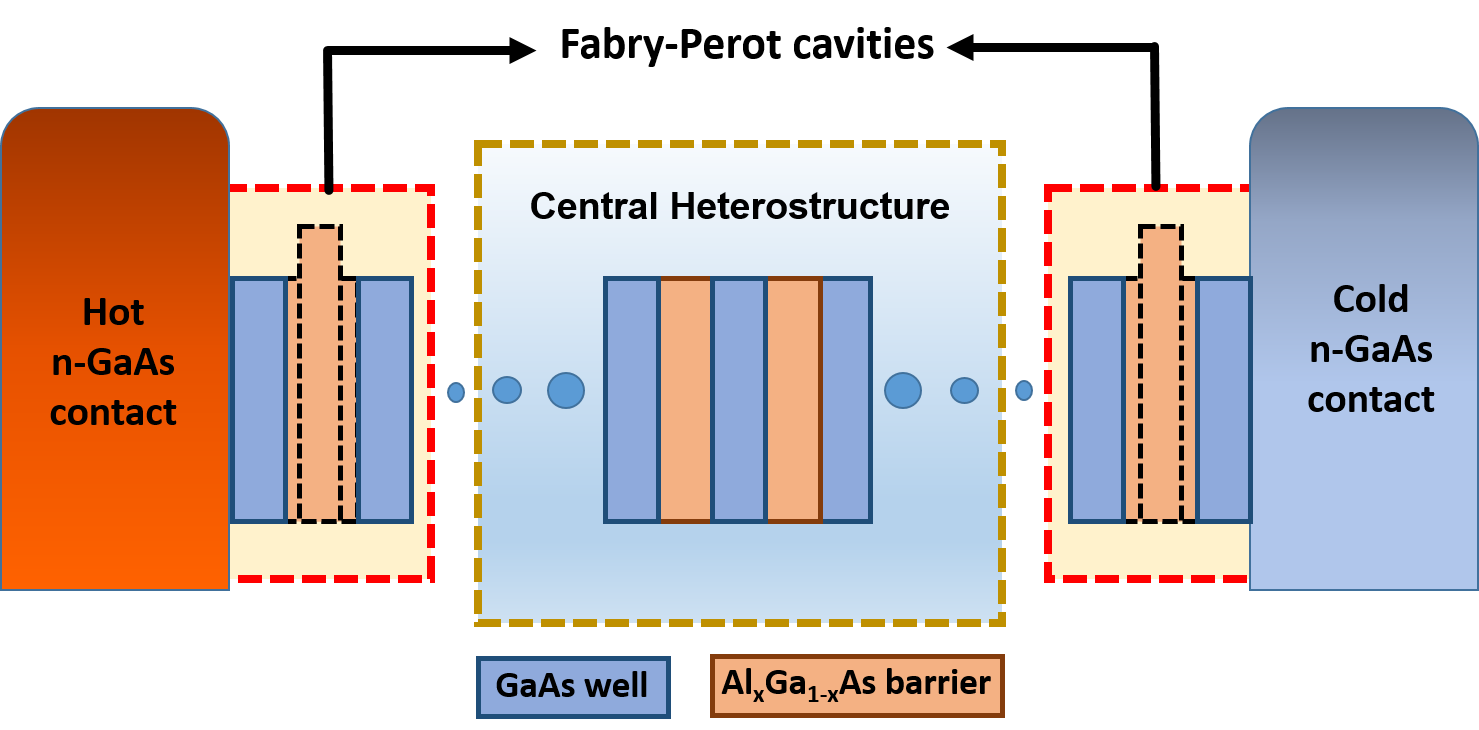}}
	\caption{Device schematic of an electronic Fabry-P\'erot cavity engineered heterostructure based thermoelectric generator setup. The central region, in general, consists of a multi period heterostructure sandwiched between two electronic cavity sections. This work considers the use of a simple double barrier resonant tunneling structure embedded by cavities of varying wall geometry to optimize the desired shape of transmission spectrum.}
	\label{devschema}
\end{figure}
However, it should be noted that, in the presence of charging effects, the superlattice designs \cite{pankaj} suffer from serious lineshape imperfections which badly affects the power and trade-off characteristics. Moreover, the large number of constituting layers in such devices poses a serious threat to the precise epitaxial growth with the existing technology. On the other hand, the ARC based proposal \cite{myTED} although produced improved result but was never optimized for further scope of improvements. The object of this paper is to hence propose a TEG device structure and explore its design space to provide a robust design guideline after examining and taking into consideration all the aforementioned aspects.

\indent In this work, we consider a simple double barrier resonant tunneling (RT) structure embedded in an electronic Fabry-P\'erot (FP) cavity as shown schematically in Fig.~\ref{devschema}. The dotted rectangle in the cavity region denotes the variation of the width and height of the electronic potential barrier. This cavity is similar to a Fabry-P\'erot setup used in optics where the mirrors are replaced by rectangular tunneling electronic barriers which act as cavity walls. The transmission function, being strongly dependent on the tunneling probability through these barriers, can be tuned by varying their height and width. We show that by following a specific design guideline, a nearly band-pass transmission can be achieved by varying the wall geometry. A careful examination of the transmission function reveals that one can achieve even wider band-pass profile compared to that of the conventional ARC based design \cite{myTED,myTF} by following the proposed guideline. This setup when used as a thermoelectric generator can significantly raise the output power at a high conversion efficiency as compared to the existing proposals \cite{Agarwal2014,pankaj,Nakpathomkun2010}. Exploring the design space further, it is seen that an improvement of output power up to 18\% can be achieved without any degradation in the efficiency over the ARC based structure \cite{myTED}. \\
\indent The rest of the paper is structured as follows. In Sec.~\ref{sec_trans}, the variation of the transmission function with respect to the different cavity designs is thoroughly examined and explained in the lights of ARC physics. Based on the obtained result, three unique designs are picked for further investigation on their capability of being good thermoelectric generator. The band schematics of all the devices are depicted in Sec.~\ref{dev} with a clear description of their physical properties. Section~\ref{simu} briefly discusses the simulation setup and illustrates the formalism used. In Sec.~\ref{res}, the results are thoroughly discussed in terms of all the performance parameters and a detailed comparative study is presented in order to highlight the improvements achieved through the proposed design scheme. We conclude the paper in Sec.~\ref{con}. 

\section{Cavity Physics and Transmission Function}
\label{sec_trans}
In this section, we closely inspect the variation of the transmission function, $T(E)$, with respect to the stoichiometric and geometric changes of the cavity wall. The thumb rule of designing ARC says that the width ($b_{FP}$) and height ($h_{FP}$) of the rectangular cavity barriers should exactly be half and equal, respectively, to that of the central barrier region \cite{Pacher2001}. The reason behind this can be qualitatively explained in terms of the  electronic Bloch states in the neighborhood of the transmission peaks of the central heterostructure \cite{Pacher2001,Martorell2004,Morozov2002,myTED}. According to the modified Kronig-Penney model, the transmission peaks of the periodic heterostructure occur when \cite{Pacher2001,Bastard}
\begin{equation}
cos(kL)=cos \left( \frac{i\pi}{N} \right) , \quad i=1,2,...,N-1,
\label{KP}
\end{equation} 
where $L$ is the length of the periodic structure, $N$ is the number of periods and $k$ is the Bloch wave vector which is defined as $k=\frac{2\pi}{\lambda}$, where $\lambda$ is the wavelength. Replacing $k$ by $\lambda$ in Eq.~\ref{KP}, we get $\lambda_{i}=2L/i$, which says that twice the length of the structure should be equal to integer multiple of the allowed wavelengths. The concept of electronic anti-reflection is actually borrowed from the well-known Fabry-P\'erot setup used in optics. To satisfy the anti-reflection condition, the reflected waves from the two boundaries of the cavity barrier should exactly be out of phase of each other. In other words, the cavities should act as Bragg-reflector at a wavelength $\lambda^{'}$ which satisfies the condition for thin film interference, given by 
\begin{equation}
2b_{FP}=\left( m+ \frac{1}{2} \right)\lambda^{'},
\label{ARC}
\end{equation}
where $\lambda^{'}$ lies in the neighborhood of $\lambda$ and $m$ is an integer. Therefore, for $m=0$, a unity transmission peak occurs at $\lambda^{'}$ if the cavity barriers are $\lambda^{'}/4$ layers ($b_{FP}=\lambda^{'}/4$). This condition along with aforementioned relation of $L=i\lambda/2$ suggest that $b_{FP}$ should be around half the width of the central barrier region ($b$) as the width of the well regions ($w$) throughout the structure are considered to be uniform. \\
In this context, one should always ponder that unlike the optical setup, the height of the cavity wall plays a crucial role on tailoring the lineshape of the transmission. To be more specific, the combined effect of $b_{FP}$ and $h_{FP}$ controls the phase of the reflected waves from the cavity wall boundaries which in turn determines the transmission probability. By carefully examining the transmission of a setup shown in Fig.~\ref{devschema}, we note that the amount of aberration from the bandpass nature caused by a tiny reduction in $b_{FP}$ from $b/2$, can be compensated by a proportional upscaling of $h_{FP}$ from $h$. To explain this, we draw a connection between the potential energy of the cavity barrier region ($h_{FP}$) and its refractive index ($n$). It should be noted that for a medium with refractive index $n$, the associated wavelength ($\lambda_n$) is defined as $\lambda_n=\lambda/n$, where $\lambda$ is the corresponding vacuum wavelength. Therefore, replacing $\lambda^{'}$ by $\lambda^{'}/n$ in Eq.~\ref{ARC} for $m=0$, the condition for anti-reflection becomes
\begin{equation}
b_{FP}=\frac{\lambda^{'}}{4n}.
\label{AR}
\end{equation}
In the wave-particle duality picture, this wavelength is called the de-Broglie wavelength of the electron which is directly related to its momentum ($p$) by the relation $\lambda=h/p$. Hence the local de-Broglie wavelength of the cavity tunnel barrier can be expressed as 
\begin{equation}
\lambda_n=\frac{\lambda}{n}=\frac{h}{\sqrt{2m(E-h_{FP})}},
\end{equation}
where $p=\sqrt{2m(E-h_{FP})}$ for a rectangular barrier of height $h_{FP}$, $m$ is the effective mass of the tunnel barrier, $h$ is the Planck's constant, E is the electron energy and $\lambda$ is the reference wavelength (here, the wavelength of the well region on both sides of the tunnel barrier). The refractive index $n$ is thus given by \cite{ARsrep,FPstolle}
\begin{equation}
n=\frac{\lambda}{\lambda_n}=\sqrt{\frac{2m(E-h_{FP})}{2m_0 E}},
\label{refrac}
\end{equation}
where the well region is having an effective mass of $m_0$ and zero potential energy. As we are only concerned in the energies below the cavity barrier height ($E<h_{FP}$), the condition for anti-reflection is thus obtained by substituting the absolute value of $n$ from Eq.~\ref{refrac} into Eq.~\ref{AR} which is given by
\begin{equation}
b_{FP}=\frac{\lambda^{'}}{4}\sqrt{\frac{2m_0 E}{2m(h_{FP}-E)}}.
\end{equation}
This indicates that for the anti-reflection condition to prevail, any reduction in $b_{FP}$ must be associated with a particular increase in $h_{FP}$. As the design energy $E$ can't be precisely defined, one can't establish a specific relation between $b_{FP}$ and $h_{FP}$. However, one can always predict an optimal design guideline by examining the transmission function of the RT structure embedded in an electronic FP cavity. \\
\begin{figure}
	\centering
	\subfigure[]{\includegraphics[scale=0.35]{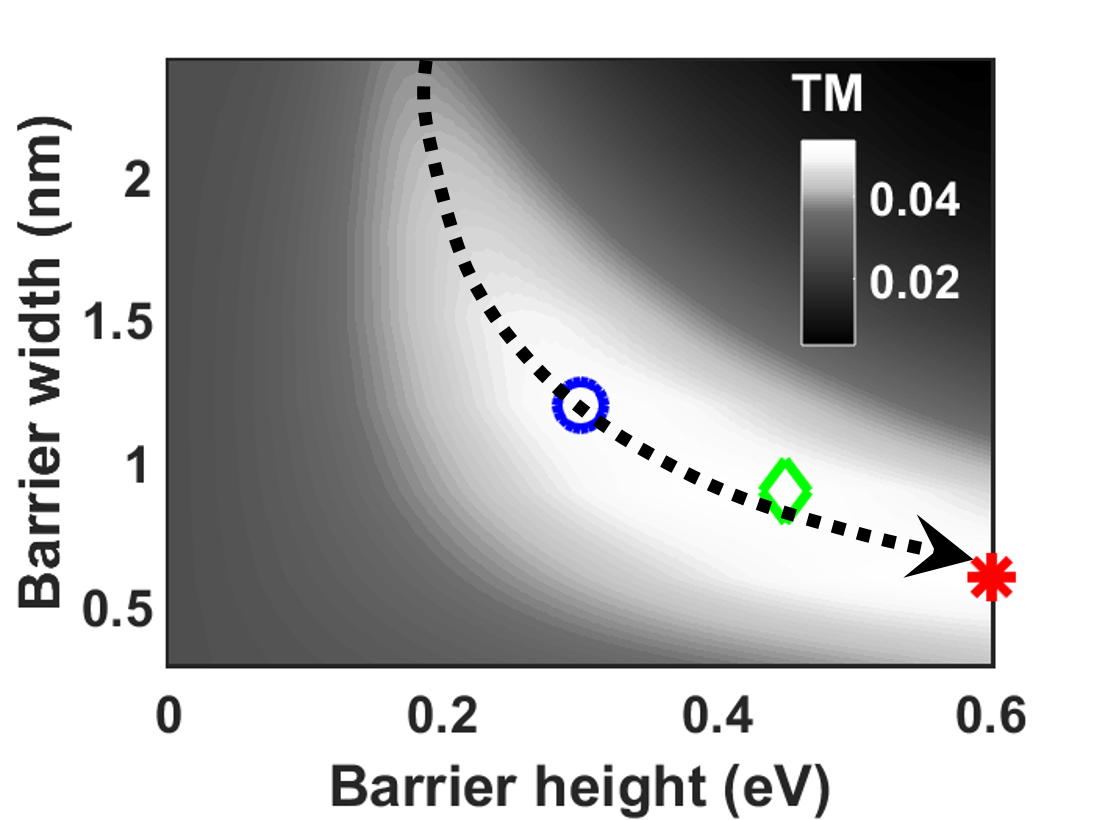}\label{trans1}}
	\subfigure[]{\includegraphics[scale=0.35]{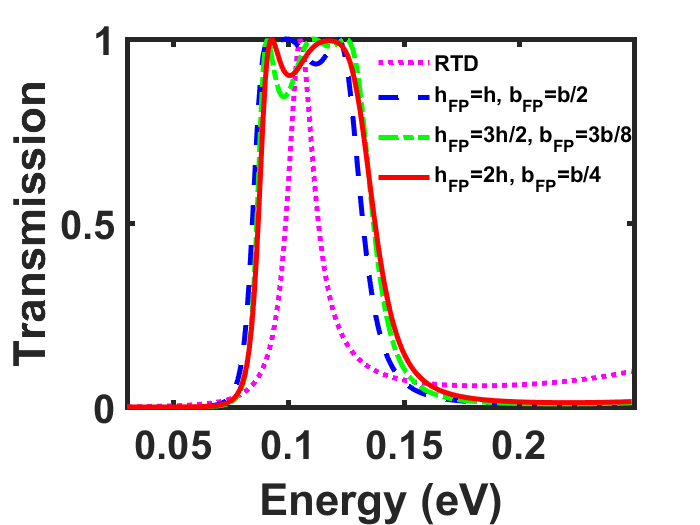}\label{trans2}}
	\caption{Transmission function: (a) Area under the flatband transmission function corresponding to the lowest transmission band (TM) is shown in a gray scale color plot as a function of the cavity wall width and height. The locus of its maxima follows a nearly hyperbolic trend which increases along the direction of the black dotted arrow. From the obtained trend, two new design schemes are picked (green and red) for further investigation as good thermoelectric generators and comparison with the ARC based proposal (blue). (b) Equilibrium flatband transmission function of all the cavity based devices are shown as a function of energy. The peaked transmission of the central RT region (without the ARC) is also shown here to emphasize the role of cavity engineering on transmission.}
	\label{trans}
\end{figure}
Above theory calls for a further investigation on the possible betterment of the boxcar nature of the transmission \cite{myTED} by means of optimal cavity engineering. A quantitative measure in this regard is the transmissivity ($TM$) which is the area under the flatband transmission function corresponding to the lowest transmission band, given by
\begin{equation}
TM=\int_{0}^{E_{1}} \left| T(E) \right| dE,
\end{equation}
where the energy $E_{1}$ is chosen in such a way that it falls almost in between the ground and first excited band with almost zero transmission probability. The transmission function is calculated using the standard non-equilibrium Green's function (NEGF) theory \cite{QTDatta} which will be addressed later. Figure \ref{trans}(a) displays the variation of $TM$ of the aforementioned setup as a function of $b_{FP}$ and $h_{FP}$ in a gray scale color plot for a given set of RT device parameters which will be discussed in the next section. We observe that $TM$ exhibits a nearly hyperbolic trend around its maxima which monotonically increases (along the direction of the black dotted arrow) with decreasing $b_{FP}$ and increasing $h_{FP}$. A careful investigation of the transmission reveals that its boxcar nature can almost be maintained if the percentage reduction in $b_{FP}$ from $b/2$ is equal to half the percentage increase in $h_{FP}$ from $h$. This finding closely matches with the theory presented above. Therefore, the design guideline to achieve boxcar transmission can be mathematically expressed as 
\begin{equation}
\frac{\left| b_{FP} - b/2 \right|}{b/2} =\frac{1}{2} \frac{\left| h_{FP} - h \right|}{h}. 
\end{equation}
The allowed design space of $b_{FP}$ and $h_{FP}$ is given by $b_{min} \leq b_{FP} \leq b/2$, $h \leq h_{FP} \leq h_{max}$, where $b_{min}$ and $h_{max}$ are the practical bounds of cavity barrier width and height, respectively. In this case, based on the desired transmission goal, these bounds are set as $b_{min}=b/4$ and $h_{max}=2h$.\\
It is also worth mentioning that within the allowed design space, the steady increase of $TM$ along the direction shown in Fig.~\ref{trans}(a) suggests that the boxcar nature can be further improved by utilizing other set of cavity designs. In order to justify this, we pick two sample design schemes of the FP cavity namely, FP-II (green diamond) and FP-III (red star) as indicated in Fig.~\ref{trans}(a) alongside the typical ARC ($b_{FP}=b/2$, $h_{FP}=h$) based proposal (FP-I, blue circle) \cite{myTED,myTF}. Under flatband conditions, the equilibrium transmission function of all the FP based designs are plotted in Fig.~\ref{trans}(b) as a function of energy along with the standard RT transmission. The cavity design parameters corresponding to all the devices are presented in the legends of Fig.~\ref{trans}(b) in terms of the RT design parameters. We observe that as compared to FP-I, the new schemes (FP-II and FP-III) tend to widen the transmission further preserving its desired shape, thereby improving $TM$. One notable difference to notice here is that the new proposals exhibit a slight dip in the transmission at energies below the resonating peak unlike the ARC based design. This might cause a slight reduction in the efficiency at lower values of the contact Fermi level. It is also important to note that the cavity region, based on its design, pulls the transmission minima to unity at a particular energy which might not be the mid-band energy always. In this case, Fig.~\ref{trans}(b) suggests that as the width of the cavity barrier is reduced, this energy tends to rise which in turn widens the transmission bandwidth. Having obtained such transmission features, we strongly believe that the new designs can be even better thermoelectric generator and hence should be investigated further.
\section{Device Schematic and Description}
\label{dev}
\begin{figure}
	\subfigure[]{\includegraphics[height=0.14\textwidth,width=0.4\textwidth]{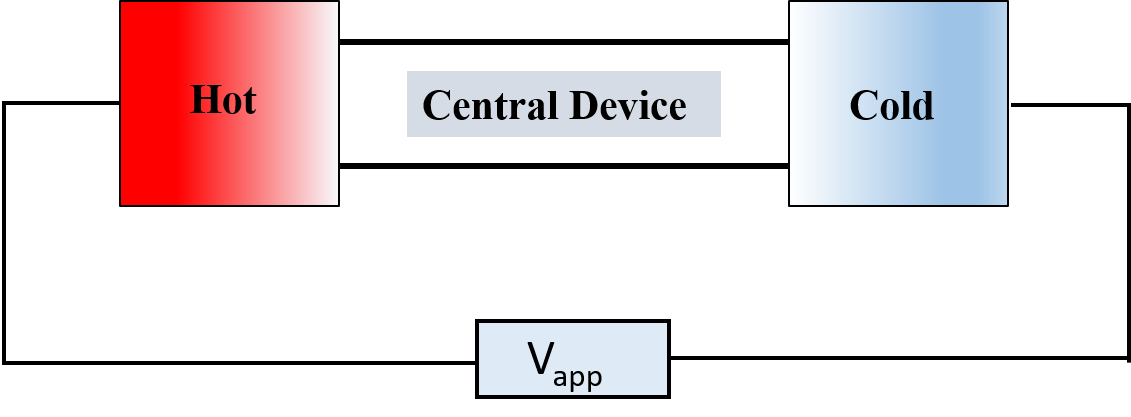}\label{schemaFP1}}
	\quad
	\subfigure[]{\includegraphics[height=0.1\textwidth,width=0.4\textwidth]{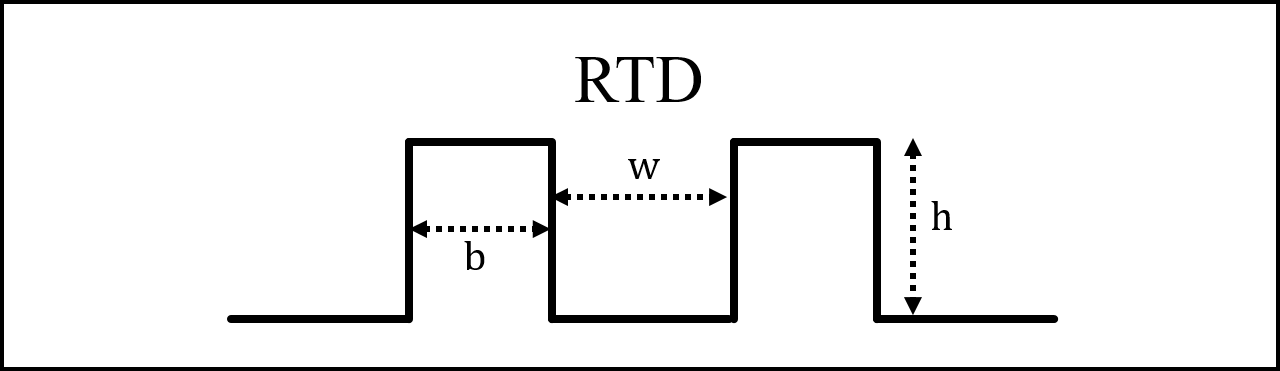}\label{schemaFP2}}
	\quad
	\subfigure[]{\includegraphics[height=0.1\textwidth,width=0.4\textwidth]{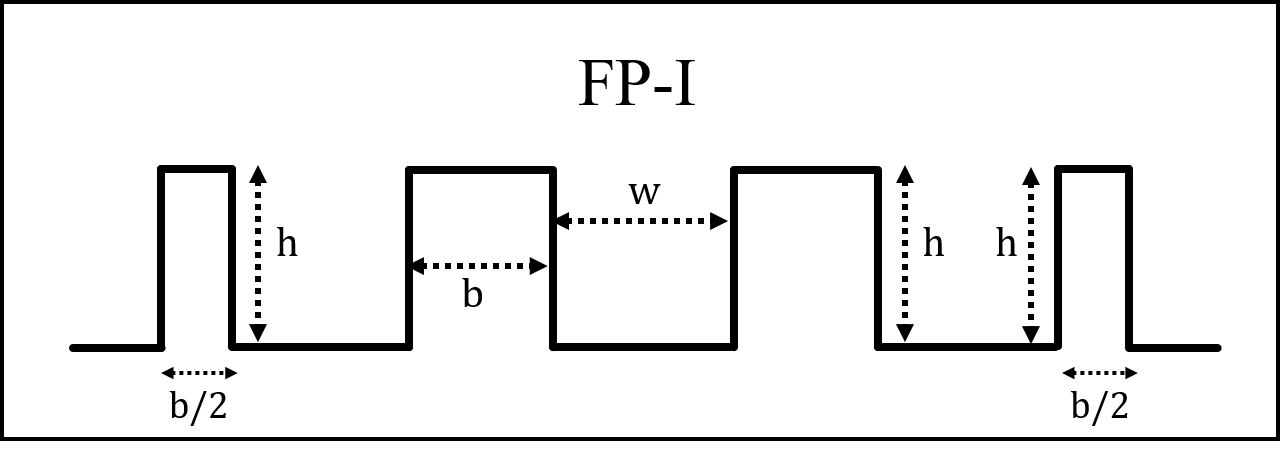}\label{schemaFP3}}
	\quad
	\subfigure[]{\includegraphics[height=0.11\textwidth,width=0.4\textwidth]{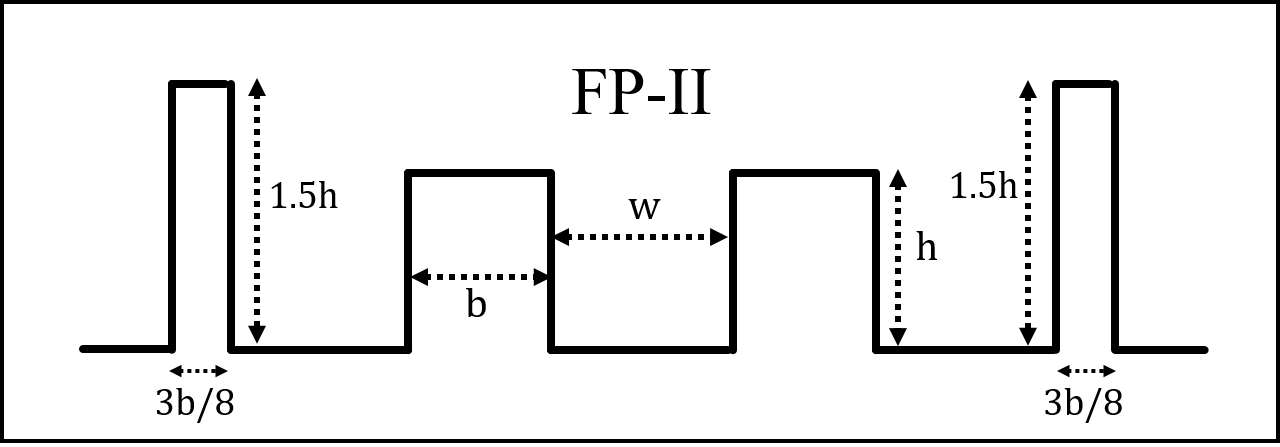}\label{schemaFP4}}
	\quad
	\subfigure[]{\includegraphics[height=0.12\textwidth,width=0.4\textwidth]{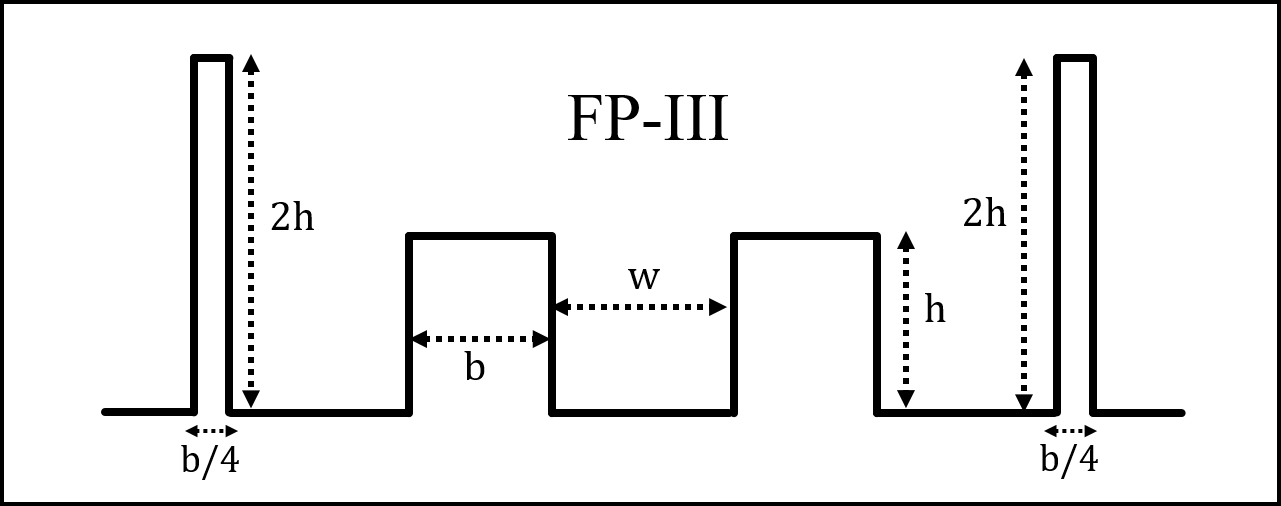}\label{schemaFP5}}
	
	\caption{Simulation setup and device schematics: (a) A typical voltage-controlled thermoelectric setup is shown where the central device is connected to two contacts of different temperatures externally joined by a load. The conduction band schematics of four different TE device structures are depicted as follows: (b) RTD-TE: a standard resonant tunneling structure having the transmission $FWHM=k_{B}T/2$. This RTD device is embedded into three different FP cavity configurations namely, (c) FP-I: $h_{FP}=h$ and $b_{FP}=b/2$, (d) FP-II: $h_{FP}=3h/2$ and $b_{FP}=3b/8$ and (e) FP-III: $h_{FP}=2h$ and $b_{FP}=b/4$.} 
	\label{schemaFP}
\end{figure}

\indent Based on the design rules discussed in the last section, we depict the conduction band schematics of all the three cavity engineered devices (FP-I, FP-II and FP-III) along with the standard resonant tunneling device (RTD) in Fig.~\ref{dev}(b)-(e). These devices are having an ideal infinite extent in the transverse direction with a finite length along their transport direction (here, z-direction). The central RTD structure, as shown in Fig.~\ref{schemaFP}(b), is modeled with a GaAs well of width $w=4.2 nm$ in between two $Al_{x}Ga_{1-x}As$ barriers of width $b=2.4 nm$ each, where $x$ is aluminum mole-fraction. Barrier height is kept fixed at $0.3 eV$ with respect to the well by precisely tuning the mole-fraction parameter. These design parameters are chosen in accordance with a realistic ground state transmission full width at half maximum (FWHM) of $k_{B}T/2$, where $k_B$ is the Boltzmann's constant and $T$ denotes the temperature. For the cavity based devices, the same RTD structure is symmetrically placed within the cavity regions such that the width of the well region between any two successive barriers remains the same at $w$. However, the varying design of cavity wall gives rise to three different structures considered in this study which are listed below:
\begin{itemize}
	\item In Fig.~\ref{schemaFP}(c), FP-I: $h_{FP}=h$ and $b_{FP}=b/2$,
	\item In Fig.~\ref{schemaFP}(d), FP-II: $h_{FP}=3h/2$ and $b_{FP}=3b/8$,
	\item In Fig.~\ref{schemaFP}(e), FP-III: $h_{FP}=2h$ and $b_{FP}=b/4$.
\end{itemize}  
\indent The devices described above can be fairly accurately modeled using a nearest neighbor tight-binding Hamiltonian of a linear atomic chain within the single-band effective mass approximation \cite{QTDatta}. The GaAs/AlGaAs material system is chosen here due to its less variability of effective mass over a wide range of composition and excellent lattice matching capability. Using the NEGF technique coupled with the charging effect, we present a comparative study of the devices discussed above in terms of the linear and non-linear thermoelectric performance parameters. The device dimensions used here are in the order of the relaxation length scales which eliminates the possibility of scattering to ensure a coherent transport of carriers within the ballistic limit \cite{Agarwal2014}. On the other hand, the presence of nano-structured interfaces strongly restricts the flow of phonons in the device. This implies that the heat current flowing through the device is mainly due to electrons. Therefore, the lattice contribution to the thermal conductivity is ignored here. \\
\indent The cavity based devices, manifest high immunity to the non-equilibrium changes in transmission function due to the charging effect. This results in an improved trade-off characteristics for a wide range of contact Fermi level. Furthermore, the widening of the transmission window allows a large number of additional transverse modes to conduct and contribute to the net charge current which in turn boosts the power. Based on the results, we can definitely assert that the width of the transmission function obtained here is still below the ideal theoretical limit predicted by Whitney \cite{Whitney2014} which makes a room for further research.
\section{Simulation Methodology and Setup}
\label{simu}
Figure \ref{schemaFP}(a) shows a typical voltage-controlled thermoelectric heat engine setup \cite{LNEDatta} which will be used throughout for the purpose of simulation. The flow of electrons due to the thermal driving force from the hot to cold contact is opposed by the voltage drop across the load resistance connecting them. The polarity of this drop is such that it lowers the quasi Fermi level of the hot contact with respect to the cold contact which, as a result, causes an opposite flow of electrons. In the simulation framework, the variation of the load resistance is incorporated through the application of a positive voltage at the hot contact end. \\
\indent The simulation methodology is mainly divided into two important parts, namely, (i) self-consistent estimation of the electronic transmission function and (ii) the calculation of charge and heat currents from the knowledge of the obtained transmission function. For the former part, we utilize the standard atomistic NEGF formalism \cite{LNEDatta,QTDatta} self-consistently coupled with the Poisson's equation. In order to analyze the device behavior under different operating conditions, we vary the equilibrium quasi Fermi levels ($E_f$) of the hot ($\mu_{H}$) and cold ($\mu_{C}$) contacts. For a given applied bias of $V_{app}$, the Fermi level of the hot (cold) contact is shifted downward (upward) from its equilibrium value by an amount of $qV_{app}/2$ due to symmetric electrostatic coupling, where $q$ is the unit electronic charge. The simulation begins with a linear potential profile as an initial guess to calculate the longitudinal energy ($E$) resolved retarded Green's function $G(E)$, given by
\begin{equation}
G(E)=[(E+i0^{+})\mathbb{I}-H-U(z)-\Sigma_H(E)-\Sigma_C(E)]^{-1},
\label{eqG}
\end{equation}
where $U(z)$ is the potential profile along the transport direction, $\Sigma_{H(C)}$ is the self-energy matrix of the hot (cold) contact and $\mathbb{I}$ is the identity matrix. Having obtained $G(E)$, the carrier concentration ($n$) can be easily calculated from the electron correlation function, $G^{n}(E)$, which is then fed into the Poisson's equation to calculate the updated potential profile. The set of equations governing the above mentioned routine are given by
\begin{equation}
G^n(E)=G[\Gamma_{H} f_{2D}(\mu_{H}) + \Gamma_{C} f_{2D}(\mu_{C})]G^{\dagger},
\label{eqGn}
\end{equation}
\begin{equation}
n = \frac{1}{\Delta z} \int\frac{G^n(E)}{2 \pi} dE,
\label{enz}
\end{equation}
\begin{equation}
\frac{d^2}{dz^2}(U(z)) = \frac{-q^2}{\epsilon_0 \epsilon_r} n
\label{eqPoisson}
\end{equation}
where $\Delta z$ is the discrete lattice spacing parameter, $\epsilon_0$ is the free space permittivity, $\epsilon_r$ is the relative permittivity of GaAs which is assumed to be uniform throughout the lattice and $\Gamma_{H(C)}$ represents the broadening function of hot (cold) contact which is defined as $\Gamma_{H(C)}=i\left[ \Sigma_{H(C)} - \Sigma_{H(C)}^\dagger \right]$. The contribution from all the transverse modes are encapsulated in the $f_{2D}$ function which is defined as \cite{QTDatta}
\begin{equation}
f_{2D}(E-\mu) = \frac{m_e^* k_B T} {2\pi\hbar^2} \log[1+exp(\frac {\mu-E} {k_BT})],
\label{eqF2D}
\end{equation}
where $\hbar$ is the reduced Planck's constant and $m_{e}^{*}$ is the electron effective mass which is considered to be uniform throughout the lattice. For our simulations, we take a constant effective mass of $0.07m_0$ across structures, where $m_0$ is the free electron mass. The NEGF-Poisson simulation is performed self-consistently until the convergence is achieved and the non-equilibrium transmission function, $T(E)$, can thereby calculated as 
\begin{equation}
T(E)=Tr[\Gamma_{H} G \Gamma_{C} G^\dagger].
\label{eqTM}
\end{equation}
The resultant transmission function is then fed into the Landauer current formula to calculate the charge ($J$) and heat current ($J^{Q}$) densities \cite{QTDatta}. Summing over all the current carrying transverse modes and absorbing that in the $f_{2D}$ function, total charge current flowing through the device is given by
\begin{equation}
J=\frac{q}{\pi \hbar} \int dE T(E) [f_{2D}(E-\mu_{H})-f_{2D}(E-\mu_{C})]. 
\end{equation}
It is important to note that the total heat current which is the energy weighted charge current, is resolved into two components namely, $J_{H}^{Q1}$ and $J_{H}^{Q2}$ based on the contributions from longitudinal and transverse energy degrees of freedom, respectively. Therefore, the total heat current flowing through the hot contact ($J_{H}^{Q}$) is expressed as $J_{H}^{Q}=J_{H}^{Q1}+J_{H}^{Q2}$, where $J_{H}^{Q1}$ and $J_{H}^{Q2}$ are given by
\begin{multline}
J_{H}^{Q1}=\frac{1}{\pi \hbar} \int dE T(E) (E-\mu_{H}) \\ \times [f_{2D}(E-\mu_{H})-f_{2D}(E-\mu_{C})], 
\end{multline}
\begin{equation}
J_{H}^{Q2}=\frac{1}{\pi \hbar} \int dE T(E) [g_{2D}(E-\mu_{H})-g_{2D}(E-\mu_{C})], 
\end{equation}
where $g_{2D}$ function is defined as \cite{Agarwal2014,myTED}
\begin{equation}
g_{2D}(E-\mu)=\frac{m^{\star}}{2\pi \hbar^{2}} \int_{0}^{\infty} \frac{\epsilon_{\vec{k}_{\perp}} d\epsilon_{\vec{k}_{\perp}}}{1+\exp(\frac{E+\epsilon_{\vec{k}_{\perp}}-\mu}{k_{B}T})}.
\label{g2deqn}
\end{equation}

The integration in Eq.~\ref{g2deqn} is performed numerically where the upper limit of energy is chosen high enough to include all the significant transverse modes. We assume a parabolic dispersion relation ($\epsilon_{\vec{k}_{\perp}}$) in the transverse direction and the integration over all the momentum ($\vec{k}_{\perp}$) eigenstates is carried out with a periodic boundary condition. \\
\indent Once the charge ($J$) and heat current ($J_{H}^{Q}$) densities are calculated, the output power density ($P$) and conversion efficiency ($\eta$) can be obtained using the standard thermoelectric setup \cite{LNEDatta} by the following relations
\begin{equation}
P=JV_{app},
\end{equation}
\begin{equation}
\eta = P/J_H^Q.
\end{equation}
The efficiency is usually measured as a ratio to that of the Carnot's limit ($\eta_{C}$), defined as $\eta_{C}=1-T_{C}/T_{H}$, where $T_{H(C)}$ is the temperature of the hot (cold) contact. In the simulation, a steady temperature difference of $30K$ is maintained between the contacts by setting $T_H=330K$ and $T_C=300K$. The allowed range of power restricts the device operation between short circuit ($V_{app}=0$) to open circuit ($V_{app}=V_{OC}$) condition, where $V_{OC}$ is the open circuit voltage.

\section{Results and Discussion}
\label{res}
In this section, a detailed and comparative study of the results are discussed in terms of the non-linear and linear response parameters. This study will mainly focus on the supremacy of the proposed device designs over the existing ones.
\begin{figure}
	\subfigure[]{\includegraphics[height=0.225\textwidth,width=0.225\textwidth]{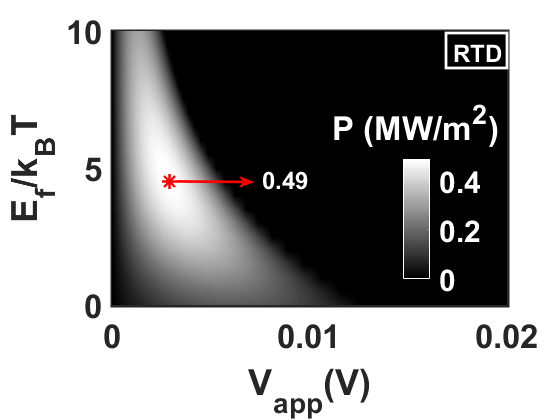}\label{pow1}}
	\quad
	\subfigure[]{\includegraphics[height=0.225\textwidth,width=0.225\textwidth]{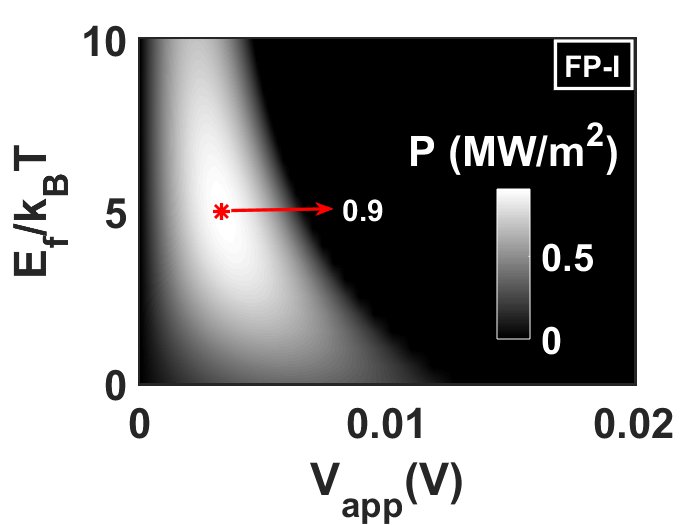}\label{pow2}}
	\quad
	\subfigure[]{\includegraphics[height=0.225\textwidth,width=0.225\textwidth]{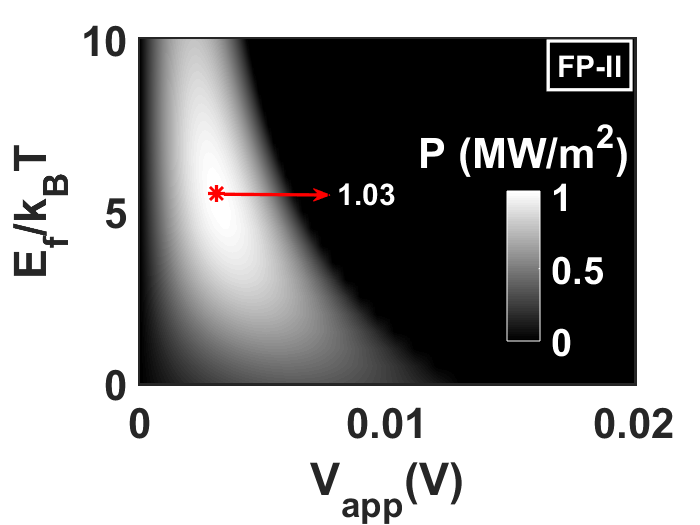}\label{pow3}}
	\quad
	\subfigure[]{\includegraphics[height=0.225\textwidth,width=0.225\textwidth]{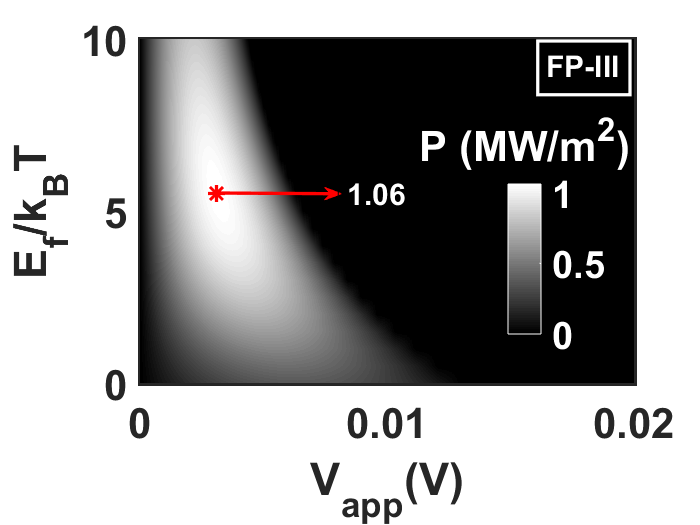}\label{pow4}}
	
	\caption{Comparative study of output power: Power density (in $MW/m^{2}$) of (a) RTD, (b) FP-I, (c) FP-II, and (d) FP-III devices are shown as a function of the applied bias ($V_{app}$) and contact Fermi level ($E_f$). Enabling ARC (FP-I) over the RTD structure nearly doubles the generated output power for the entire range of $E_f$. The power can be further boosted between $15-18\%$ by means of optimal cavity engineering as evident in case of the new design schemes (FP-II and FP-III).}
	\label{pow}
\end{figure}
\subsection{Non-linear Response Analysis}
\textbf{Power and Efficiency}: In Fig. \ref{pow}, output power per unit area of all the device structures are displayed as a function of $V_{app}$ and contact $E_{f}$ in a gray scale color plot. It can be seen that the power starts to increase from the short circuit condition with increasing $V_{app}$ and reaches a local maxima before falling to zero at the onset of the open circuit condition. Strictly speaking, net current actually reverses its direction at $V_{OC}$ and therefore the setup can't be used as a generator beyond this point. Usable power in the region beyond $V_{OC}$ is thus treated as zero. This trend is almost similar irrespective of the design scheme, but, what is important to note here is the variation of power with $E_{f}$. When $E_{f}$ is moved up in the energy scale from the lowest conduction band edge, the net flow of electrons from the hot to cold contact increases steadily. This results in a monotonic rise of power until it reaches its peak value when the net electron flow becomes maximum. At this point the overlap between the electronic density-of-states (DOS) and the region where $f_{2D}(\mu_H)-f_{2D}(\mu_C)>0$ becomes maximum which also indicates to the most non-reversible state of the heat engine. With further increase in $E_{f}$, power starts to die down steadily as the reverse flow (cold to hot) of electrons increases until $E_{f}$ moves in the vicinity of the higher excited states. But, we restrict our study only within the contribution of the ground state as the excited states hardly contribute to the conduction due to their negligible electron population and is thus kept out of consideration. On the other hand, when $E_{f}$ goes way down in energy, the power becomes negligible due to the lack of available states for conduction in the Fermi window. We, therefore, set the range of $E_{f}$ between $0-10k_{B}T$ in our simulation where the reference energy $E=0$ is chosen as the conduction band minimum of GaAs. \\

\indent Figure \ref{pow}(a) displays the power density profile of the RTD-TE device which reveals that the maximum power of $0.49 MW/m^{2}$ can be delivered at $E_{f}=4.5k_{B}T$. It is also important to observe that with increasing $E_{f}$, $V_{OC}$ sharply falls due to the sharp nature of the transmission and therefore the power remains non-zero only for a narrow region of operation. On the other hand, the cavity based devices due to their band-pass nature of transmission, manifest a huge improvement in the power along with a broad spectrum as depicted in Fig.~\ref{pow}(b), (c), (d) for the configurations FP-I, FP-II and FP-III, respectively. Obtained results show that FP-II and FP-III designs can generate maximum power ($P_{max}$) up to $1.03 MW/m^{2}$ and $1.06 MW/m^{2}$, respectively, as compared to $0.9 MW/m^{2}$ of the ARC based proposal (FP-I) and $0.46 MW/m^{2}$ of the superlattice based generators \cite{pankaj}. The position of $P_{max}$ of the new proposals is at $E_{f}=5.5k_{B}T$ which is slightly higher than that of FP-I whose $P_{max}$ occurs at $E_{f}=5k_{B}T$. This result is in good agreement with the nature of the obtained transmission functions of the new designs as they are marginally shifted upward in energy when compared to that of FP-I. One must note that deploying the new design schemes, $P_{max}$ can be boosted up to a maximum of $18\%$ and $116\%$ over the ARC and RTD based proposal, respectively.\\ 
\begin{figure}
	\centering
	\subfigure[]{\includegraphics[height=0.225\textwidth,width=0.225\textwidth]{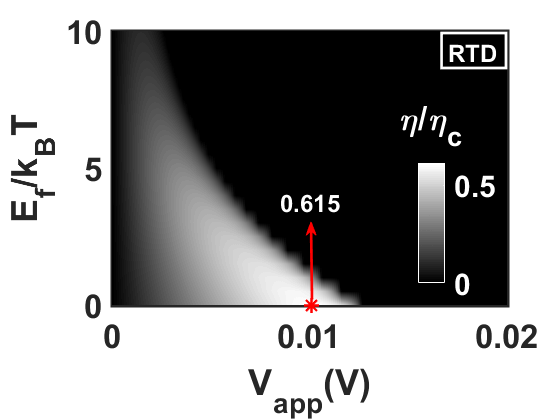}\label{eff1}}
	\quad
	\subfigure[]{\includegraphics[height=0.225\textwidth,width=0.225\textwidth]{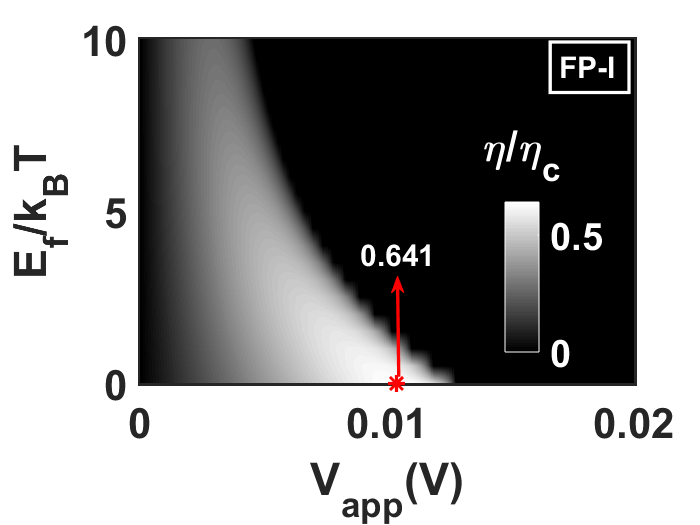}\label{eff2}}
	\quad
	\subfigure[]{\includegraphics[height=0.225\textwidth,width=0.225\textwidth]{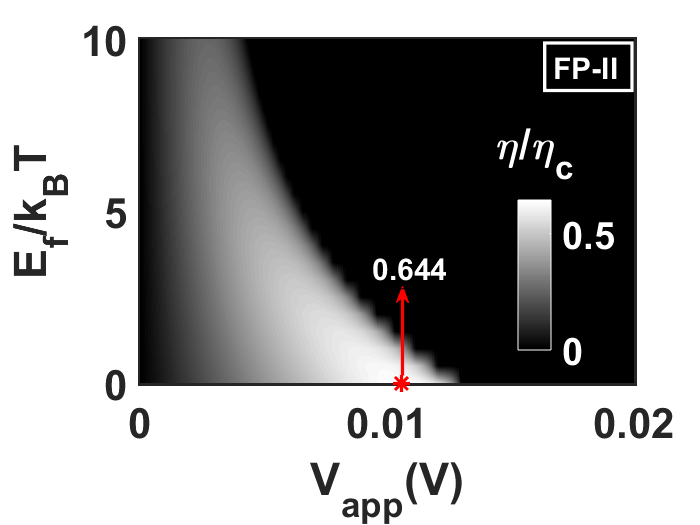}\label{eff3}}
	\quad
	\subfigure[]{\includegraphics[height=0.225\textwidth,width=0.225\textwidth]{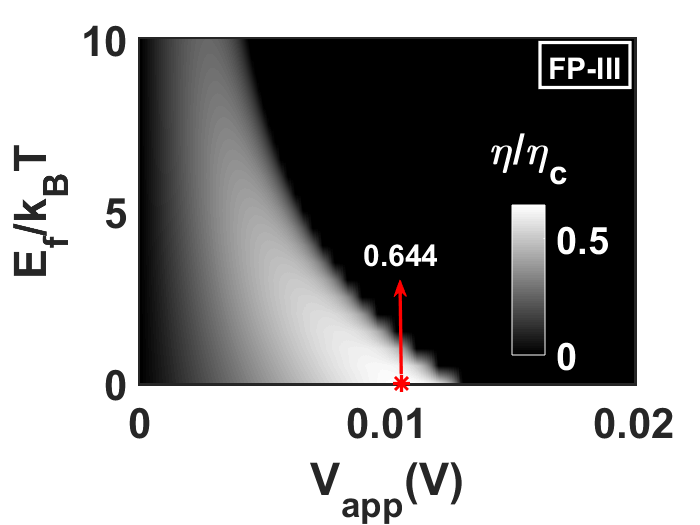}\label{eff4}}
	
	\caption{Comparative study of efficiency: Conversion efficiency normalized to Carnot's efficiency of (a) RTD, (b) FP-I, (c) FP-II, and (d) FP-III devices are shown  as a function of $V_{app}$ and contact $E_f$. The efficiency in general becomes maximum in the close proximity of $V_{OC}$ at $E_{f}=0k_{B}T$. The cavity based new proposals show almost similar range of efficiency with a hint of improvement in the maximum value as compared to the ARC based device.} 
	\label{eff}
\end{figure}
\indent A device can only be qualified as a good heat engine if it can deliver considerable amount of power at a high conversion efficiency. Therefore, an important parameter to judge here is the conversion efficiency which dictates the ability of a generator to convert heat into electricity. Normalized conversion efficiency ($\eta/\eta_C$) of all the devices are shown in Fig.~\ref{eff} as a function of $V_{app}$ and contact $E_f$. It is seen that the efficiency becomes maximum in the close vicinity of $V_{OC}$ at $E_{f}=0k_{B}T$ irrespective of the design scheme and decreases monotonically afterwards with increasing $E_{f}$. However, theoretically the efficiency can be improved further towards the ideal Carnot's limit at the cost of generated power by pushing $E_{f}$ way down the conduction band edge. But those devices would hardly be of any practical use due to their poor load driving capability. Ideally, the heat current increases when the conduction takes place at higher energies. Therefore, the efficiency attains its maximum value when $E_{f}$ is farthest below the ground transmission band. Within the mentioned simulation range, the highest efficiency that can be achieved in the RTD-TE device is $61.5\%$ at $E_{f}=0k_{B}T$ as shown in Fig. \ref{eff}(a). On the other hand, the cavity based devices although possessing wide transmission spectra, can offer even better efficiency due to their sharp transition profile of transmission as evident from Fig. \ref{eff}(b), (c), (d) for FP-I, FP-II and FP-III, respectively. The maximum attainable limit of efficiency that can be achieved through optimal cavity engineering is $64.4\%$ for the aforementioned range of power which is even better than $61.7\%$ of the superlattice based generators \cite{pankaj}. Obtained results clearly point towards an improved power-efficiency trade-off characteristics which will be discussed next.\\
\begin{figure}
	\subfigure[]{\includegraphics[height=0.225\textwidth,width=0.225\textwidth]{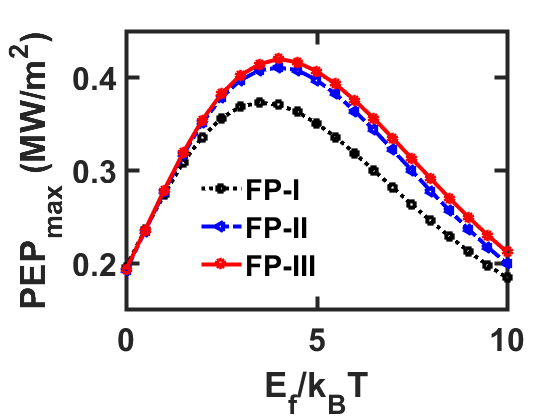}\label{pep1}}
	\quad
	\subfigure[]{\includegraphics[height=0.225\textwidth,width=0.225\textwidth]{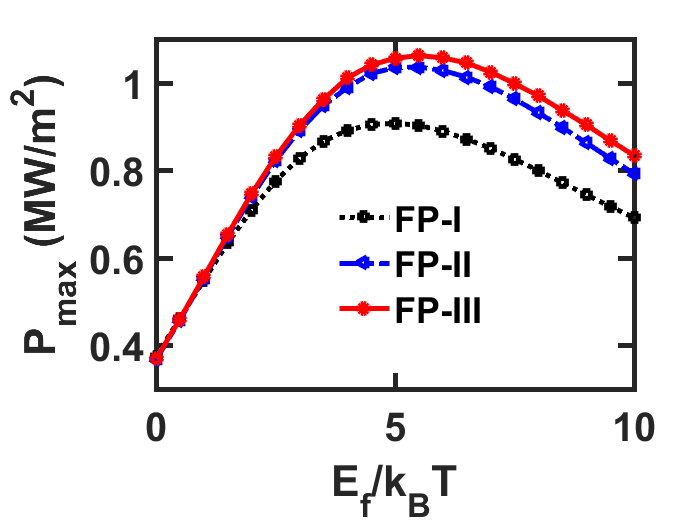}\label{pep2}}
	
	\caption{Comparative analysis: (a) $PEP_{max}$ and (b) $P_{max}$ are plotted with respect to different $E_{f}$ for all the cavity engineered devices. The difference in the range of $E_f$ pertaining to the maximum values of $PEP_{max}$ and $P_{max}$ directly point towards the trade-off between power and efficiency. It is also worth mentioning that as we move forward in the design order as in Fig. \ref{trans}(a), we achieve even more improved power and $PEP$.}
	\label{pep}
\end{figure}
\textbf{Power-efficiency-product and Trade-off}: So far, we have quantitatively discussed about the maximum achievable limit of the power and efficiency and their region of occurrence. We note that the variational trends followed by them are completely different in nature. But to design an efficient heat engine, one must be extremely careful in choosing the regime of operation such that the device can deliver significant amount of power at a high efficiency. In this context, instead of looking into the power and efficiency separately, their product ($PEP$) becomes more meaningful to inspect. For each value of $E_{f}$, the maximum of $PEP$ ($PEP_{max}$) with respect to the applied voltage is shown as a function of $E_{f}$ in Fig.~\ref{pep}(a). Besides, we also plot the maximum power ($P_{max}$) with respect to $E_{f}$ in Fig.~\ref{pep}(b) in order to compare with $PEP_{max}$. We notice that the maximum of $PEP_{max}$ occurs around $E_{f}=4k_{B}T$ which is well ahead to that of $P_{max}$ which becomes maximum around $E_{f}=5.5k_{B}T$. This clearly signifies that the efficiency falls rapidly with increasing $E_{f}$ which is also evident from the sharp fall of $PEP_{max}$ beyond its maxima in contrast to $P_{max}$. It is also worth mentioning that the margin of improvement in both the parameters becomes maximum around their respective maxima which further improves the trade-off. \\
\indent  Non-linear studies of thermoelectric heat engine has got precedence as they generally talk about the power-efficiency trade-off and the best operating regime of the device. Neither the power nor the efficiency is sufficient alone to judge the overall performance as they are dependent on each other. Therefore, we shift our attention towards determining the most suitable operating regime of these devices based on the specific design goals. A typical power-efficiency trade-off curve looks like a loop with the start (short-circuit condition) and end (open-circuit condition) points being the origin as shown in Fig. 3(a) in Ref. 30. At any particular $E_f$, the loop is obtained by plotting the efficiency against power for all values of $V_{app}$. For any loop, one can always see that $P_{max}$ and $PEP_{max}$ occur at different values of $V_{app}$. Considering both the aspects, we plot the trade-off boundaries along the locus of $P_{max}$ and $PEP_{max}$ for the series of loops at different values $E_f$ in Fig.~\ref{tradeoffFP}(a) and (b), respectively. The plots show that the trade-off characteristics improve significantly (enclosing a larger area) for the FP-II and FP-III structures as compared to the ARC based (FP-I) device \cite{myTED}. In this case, by improving we mean that the proposed devices can operate over a wide range of design parameters with satisfactory performance. A steady improvement in the trade-off begins to show up when $E_f$ goes past $3k_BT$ and maximizes in the range of $4-7k_{B}T$ for both the cases. For a given range of efficiency between $30-40\%$, the respective power (in $MW/m^2$) corresponding to $P_{max}$ and $PEP_{max}$ varies between $0.8-0.9$ and $0.75-0.87$ for FP-I, $0.94-1.03$ and $0.9-1$ for FP-II and $0.95-1.06$ and $0.94-1.03$ for FP-III. These results clearly indicate that the new proposals offer excellent trade-off characteristics and perform significantly well within the suitable operating regime of $E_f$ between $4-7k_{B}T$.
\begin{figure}
	\subfigure[]{\includegraphics[height=0.225\textwidth,width=0.225\textwidth]{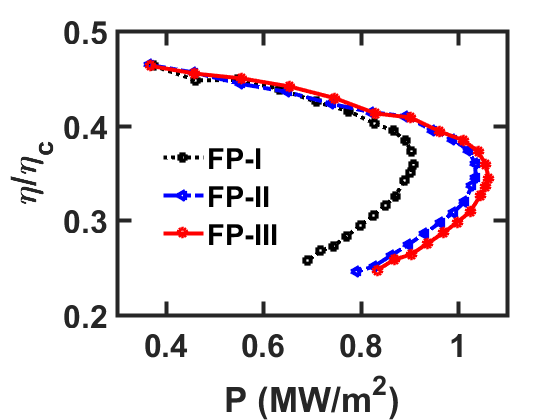}\label{pep1}}
	\quad
	\subfigure[]{\includegraphics[height=0.225\textwidth,width=0.225\textwidth]{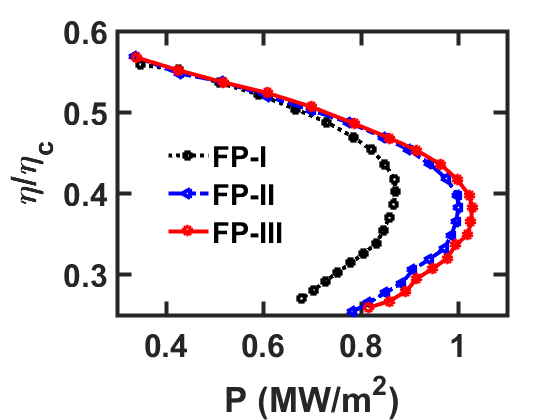}\label{pep2}}
	
	\caption{Comparative analysis of power-efficiency trade-off along the locus of (a) $P_{max}$ and (b) $PEP_{max}$ for all the cavity engineered devices. It is noted that in both the cases the new design schemes enclose a larger area on the power-efficiency plane which allows them to operate satisfactorily over a wide range of $E_f$.}
	\label{tradeoffFP}
\end{figure} 
\subsection{Linear Response Analysis}
\begin{figure}
	\subfigure[]{\includegraphics[height=0.225\textwidth,width=0.225\textwidth]{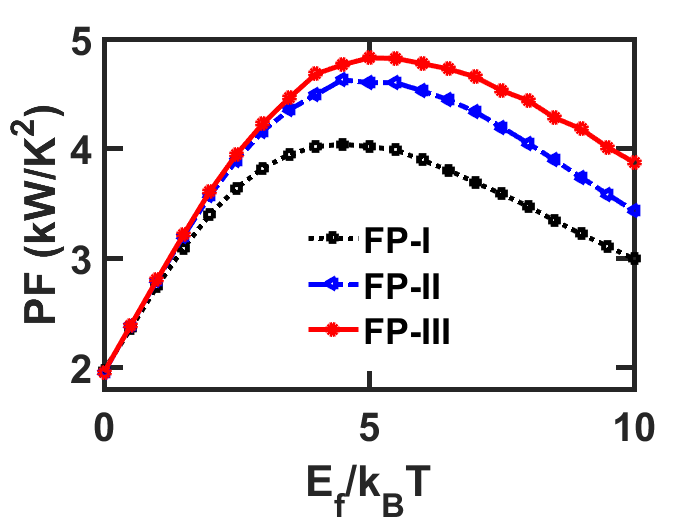}\label{pfseebeck1}}
	\quad
	\subfigure[]{\includegraphics[height=0.225\textwidth,width=0.225\textwidth]{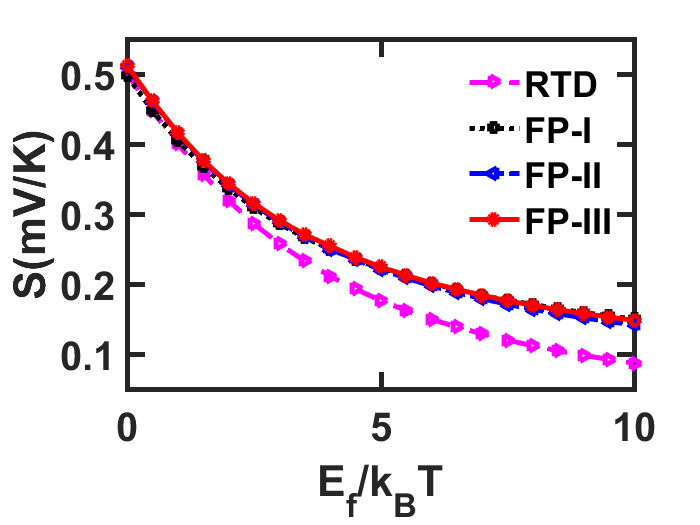}\label{pfseebeck2}}
	
	\caption{Comparative analysis: (a) Power factor ($PF$) and (b) Seebeck coefficient ($S$) are plotted with respect to varying contact $E_f$. A steady improvement in $PF$ is observed beyond $E_f=2k_BT$ as we move up in the design order from FP-I to FP-III. The maximum improvement achieved in $PF$ through cavity engineering is nearly $20\%$ at $E_f$ around $5$ to $6k_{B}T$. On the other hand, there is no noticeable difference observed in $S$ among the cavity engineered devices. However, when compared with the RTD-TE device, they show serious improvement in $S$ at higher values of $E_f$.}
	\label{pfseebeck}
\end{figure}

\begin{figure}
	\subfigure[]{\includegraphics[height=0.225\textwidth,width=0.225\textwidth]{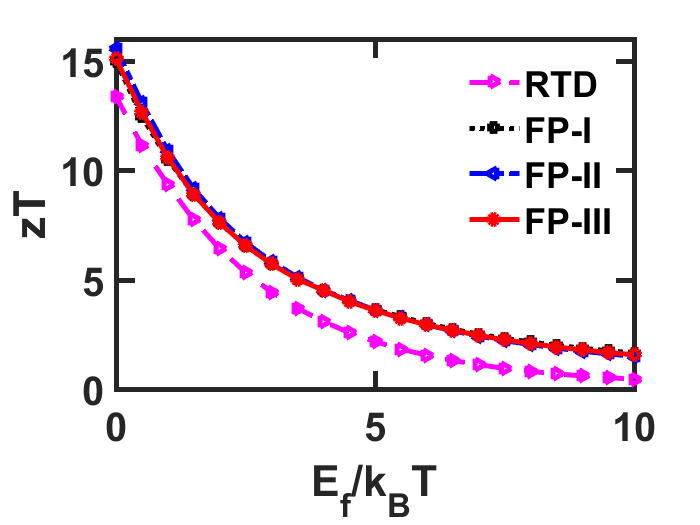}\label{zTeff1}}
	\quad
	\subfigure[]{\includegraphics[height=0.225\textwidth,width=0.225\textwidth]{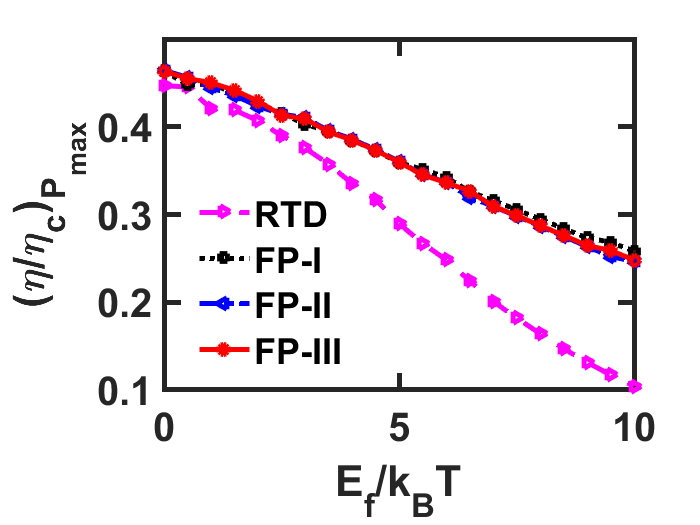}\label{zTeff2}}
	
	\caption{Comparative analysis: (a) Figure-of-Merit ($zT$) and (b) efficiency at maximum power ($\eta_{P_{max}}$) are plotted for all the devices as a function of varying $E_{f}$. The cavity based devices exhibit an almost similar $ZT$ and $\eta_{P_{max}}$ for the entire range of $E_{f}$ with a significant improvement over the RTD-TE device. The range of $zT$ and $\eta_{P_{max}}$ of the cavity based devices vary between $2.5-4.5$ and $31-39\%$, respectively, within the best operating regime of $E_f$ between $4-7k_{B}T$.}
	\label{zTeff}
\end{figure}

Using the same simulation framework, the linear response parameters can be extracted from the coupled charge and heat current equations, given by
\begin{equation}
I=G\Delta V + G_S\Delta T,\quad   I_Q = G_P\Delta V + G_Q\Delta T,
\label{eqI-IQ}
\end{equation}
where, $G$, $G_S$, $G_P$, $G_Q$ are related to the corresponding Onsager coefficients \cite{LNEDatta}. $\Delta V$ and $\Delta T$ are the applied electrical and thermal bias, respectively, which are kept small enough to ensure linear operation. \\
\textbf{Power Factor and Seebeck Coefficient}: Power factor ($PF$) is defined as $PF=S^{2}G$, where $G$ is the electrical conductivity and $S$ is the Seebeck coefficient which is given by, $S=-G_S/G$. In Fig.~\ref{pfseebeck}(a), one can easily notice the sharp and steady rise of $PF$ beyond $E_{f}=2k_{B}T$ from FP-I to FP-III. The maximum improvement in $PF$ that can be achieved through optimal cavity engineering over that of the ARC based design is nearly $20\%$ in the range of $E_f$ between $5-6k_{B}T$. This result actually points towards a monotonic improvement of $G$ as the Seebeck coefficients of the cavity based devices remain almost same for the entire range of $E_f$ as depicted in Fig.~\ref{pfseebeck}(b). We understand that the marginal improvement in the transmission function although does not affect the $V_{OC}$ much, but accounts for considerable gain in the $PF$ due to the additional large number of transverse current carrying modes that participate in conduction. \\
\textbf{Figure-of-Merit}: Although the main goal of this work is to improve the non-linear performance, however, it is customary to discuss the dimensionless Figure-of-Merit ($zT$) in order to judge the device ability as an efficient heat engine. In our study, we restrict ourselves to the electronic part of heat conduction neglecting the phonon contribution. The presence of nano-structured interfaces strongly hinders the phonon transport through the lattice which in turn results in a negligible thermal conductivity in contrast to its electronic counterpart. With these assumptions, $zT$ can be expressed as
\begin{equation}
zT = \frac{PF}{G_{K,el}} T,
\end{equation}
where $G_{K,el}$ is the open circuit electronic thermal conductivity, given by $G_{K,el} = G_Q - G_P G_S/G$. Figure \ref{zTeff}(a) plots the $zT$ of all the devices as a function of $E_f$ which clearly reveals that the boxcar feature of the transmission significantly enhances the $zT$ throughout when compared to its peaked nature. This result is also in line with the variation of efficiency at maximum power ($\eta_{P_{max}}$) with $E_f$ as depicted in Fig.~\ref{zTeff}(b). It is observed that in the cavity based devices, the achievable limit of $zT$ and $\eta_{P_{max}}$ within the suitable operating range of $E_f \simeq 4-7k_{B}T$ vary in between $2.5-4.5$ and $31-39\%$, respectively, which is pretty high as compared to $1.1-3.1$ and $20-34\%$ of a RTD-TE device. These results show that at respective maximum output power as shown in Fig.~\ref{pow}, the cavity based generators can operate at up to $10\%$ higher efficiency than that of a RTD-TE. One must also note that the range of $zT$ is almost similar in all the cavity based devices which dictates that the heat conversion ability does not degrade with an associated rise in output power. A close look on the obtained result reveals that the steady improvement of $PF$ from FP-I to FP-III is mostly suppressed by an equal rate of increase in the thermal conductivity, thereby maintaining a uniform $zT$. These results prove that the cavity engineered devices perform way better in terms of efficient heat conversion ability as compared to RTD \cite{Agarwal2014} or QD \cite{Muralidharan2012} based generators.\\
\indent The results discussed above are quantitatively summarized in Table \ref{table1} for a detailed comparative study of all the devices. This study would also help in designing suitable TE heat engines according to the specific output goals.

\setlength{\tabcolsep}{9pt}
\begin{table*}[htbp]
	
	\caption{Comparative study of key performance parameters.}
	\centering
	\begin{tabular}{|c|c|c|c|c|}
		\hline
		\textbf{Device Configuration} & \textbf{RTD} & \textbf{FP-I} & \textbf{FP-II} & \textbf{FP-III} \\
		\hline
		$P_{max}(MW/m^2)$ & 0.49 & 0.90 & 1.03 & 1.06\\
		\hline
		$\eta_{P_{max}}(\%)$ & 44.66 & 46.34 & 46.42 & 46.32\\
		\hline
		$\eta_{max}(\%)$ & 61.5 & 64.1 & 64.4 & 64.4\\
		\hline
		$PEP_{max}(MW/m^{2})$ & 0.18 & 0.37 & 0.41 & 0.42\\
		\hline
		$PF_{max}$ & 2.20 & 4.03 & 4.62 & 4.82\\
		\hline
		$zT_{max}$ & 13.37 & 14.98 & 15.57 & 15.09\\
		\hline
		$zT_{E_f\simeq 4-7k_{B}T}$ & 1.54-3.08 & 2.99-4.49 & 2.92-4.51 & 2.93-4.51\\
		\hline
		$S_{E_f\simeq 4-7k_{B}T}(mV/K)$ & 0.15-0.21 & 0.2-0.25 & 0.2-0.25 & 0.2-0.26\\ 
		\hline 
		
	\end{tabular}
	\label{table1}
\end{table*}
 
\section{Conclusion}
\label{con}
In conclusion, we have vastly explored the different design features of the electronic Fabry-P\'erot cavity over the RTD structure on achieving a nearly perfect bandpass electronic transmission. We show that there exists a specific cavity design guideline in such setups to achieve a boxcar type transmission. Based on the obtained transmission profile, we pick two sample design proposals from the allowed design space with a foresight to achieve even better thermoelectric performance than the QD, RTD, ARC or superlattice based similar existing proposals. Using the NEGF-Poisson formalism, we have presented a detailed and comparative study of the linear and non-linear performance parameters in order to justify the superiority of the cavity engineered proposals. Obtained results reveal that by following the design guideline, net deliverable power can be improved up to $18\%$ from the ARC based proposal at the same efficiency leading to an excellent trade-off between them. It is also shown that by means of cavity engineering one can achieve a maximum of $116\%$ more power at a $10\%$ higher efficiency over the RTD based heat engines. Besides, in the linear response regime, the steady improvement of the power factor does not lead to a consequent degradation in the Figure-of-merit and the Seebeck coefficient. Furthermore, we have also discussed the suitable operating regime of these devices based on the margin of improvement and specific design criteria. We believe that our study opens up a new avenue on designing transmission lineshape engineered solid state devices for various applications with the simplest of structures that can be fabricated within the existing technological framework.

{\it{Acknowlegements:}} The authors acknowledge funding from Indian Space Research Organization as a part of the RESPOND grant. This work is an outcome of the Research and
Development work undertaken in the project under the
Visvesvaraya PhD Scheme of Ministry of Electronics and
Information Technology, Government of India, being implemented
by Digital India Corporation (formerly Media
Lab Asia).

\bibliographystyle{apsrev}

\bibliography{Reference}

\end{document}